\documentclass[]{spie}  %>>> use for US letter paper
\usepackage{amsmath,amsfonts,amssymb}
\usepackage{graphicx}
\usepackage[colorlinks=true, allcolors=blue]{hyperref}
\usepackage{subcaption}

\title{Automated Classification of Body MRI Sequence Type \\ Using Convolutional Neural Networks}

\author{Kimberly Helm, Tejas Sudharshan Mathai, Boah Kim, \\Pritam Mukherjee, Jianfei Liu, and Ronald M. Summers}
\affil{Imaging Biomarkers and Computer-Aided Diagnosis Laboratory, Department of Radiology and Imaging Sciences, Clinical Center, National Institutes of Health, Bethesda, MD, USA}

\graphicspath{ 
{Figures/fig_money/} 
{Figures/fig_results/}
}

\begin{document} 
\maketitle

\begin{abstract}

Multi-parametric MRI of the body is routinely acquired for the identification of abnormalities and diagnosis of diseases. However, a standard naming convention for the MRI protocols and associated sequences does not exist due to wide variations in imaging practice at institutions and myriad MRI scanners from various manufacturers being used for imaging. The intensity distributions of MRI sequences differ widely as a result, and there also exists information conflicts related to the sequence type in the DICOM headers. At present, clinician oversight is necessary to ensure that the correct sequence is being read and used for diagnosis. This poses a challenge when specific series need to be considered for building a cohort for a large clinical study or for developing AI algorithms. In order to reduce clinician oversight and ensure the validity of the DICOM headers, we propose an automated method to classify the 3D MRI sequence acquired at the levels of the chest, abdomen, and pelvis. In our pilot work, our 3D DenseNet-121 model achieved an $F_{1}$ score of 99.5\% at differentiating 5 common MRI sequences obtained by three Siemens scanners (Aera, Verio, Biograph mMR). To the best of our knowledge, we are the first to develop an automated method for the 3D classification of MRI sequences in the chest, abdomen, and pelvis, and our work has outperformed the previous state-of-the-art MRI series classifiers.

\end{abstract}


\section{INTRODUCTION}
\label{sec:intro}  % \label{} allows reference to this section

MRI scans are widely obtained to aid in the diagnosis and treatment of diseases, such as lymphoma, breast cancer, and prostate cancer. In an MRI protocol, many sequences are obtained including T1-weighted, dynamic contrast-enhanced (DCE) series, T2-weighted, Diffusion Weighted Imaging (DWI), and derived Apparent Diffusion Coefficient (ADC) maps. Certain MRI sequences are often read in conjunction by a radiologist since they provide complimentary information related to disease status. For example, both T2-weighted MRI and Diffusion Weighted Imaging (DWI) are commonly used for diagnosing lymphoma. During the acquisition of these sequences, relevant information related to the patient and image acquisition are stored in the DICOM header. Namely, ``Body Part Examined'', ``Procedure Step Description'', ``Series Description'', and ``Protocol Name'' are common DICOM header fields. As described in prior work \cite{Anand2023_CTClass}, informative descriptions of the sequence type are seldom available, and there are many instances where conflicts might arise. For instance, the ``Body Part Examined'' field may indicate ``MRI Brain'' while the ``Procedure Step Description'' specifies ``MRI Abdomen''. Such discrepancies may arise during a busy clinical day when the DICOM header information has preset defaults by the scanner protocol and the technologists do not have adequate time to appropriately change every field in the DICOM header to reflect the patient study. Instead, they may simply edit the necessary DICOM fields. 

This is further exacerbated by the variety of MRI protocols and the parametric sequences obtained at different institutions across the world. Moreover, multiple MRI scanners from different manufacturers are also used for imaging leading to extensive differences in voxel intensity distributions across sequences as shown in Fig.~\ref{fig:mri_sequence_type}. As such, it is incumbent upon the reading radiologists and referring physicians to ``see-through'' these inconsistencies while conducting a thorough evaluation of the disease status. Such contradictions can also lead to conflicts in the radiologists' hanging protocol in PACS, requiring correction via manual intervention. In particular, they pose a significant hurdle to overcome when building a large-scale clinical cohort of patients who have undergone a specific type of MRI imaging (e.g., the MRI-GENIE study by Schirmer et al.\cite{Schirmer_majorMRIstudy}), or for the development of artificial intelligence (AI) algorithms.

\begin{figure}[!htb]
\centering
\includegraphics[width=0.95\textwidth]{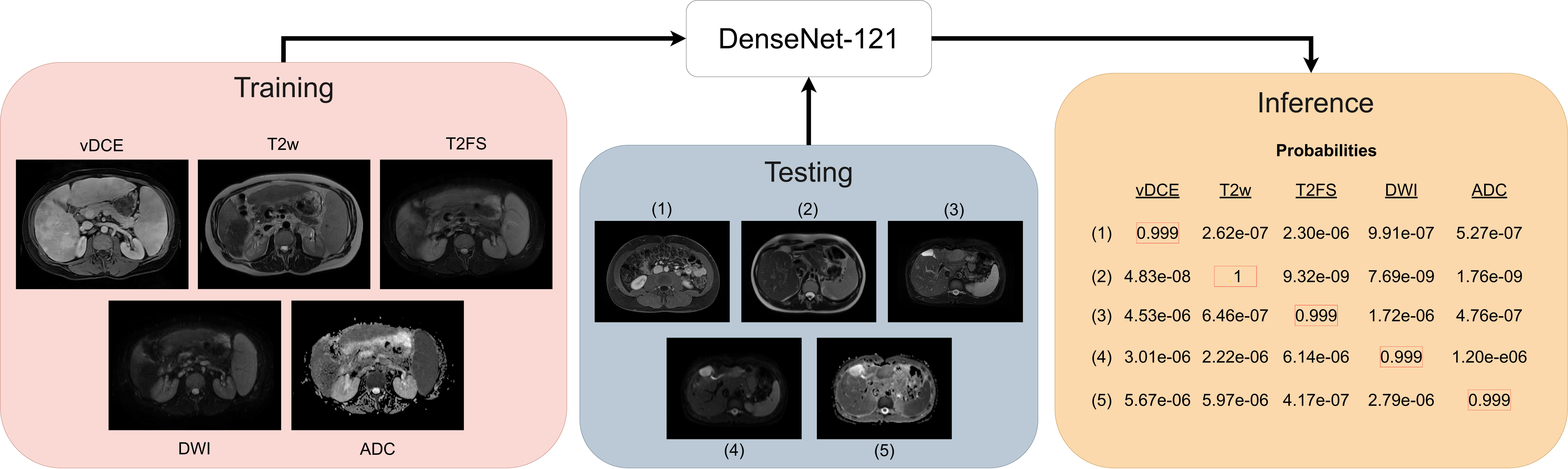}
\smallskip
\caption{Framework for the classification of multi-parametric MRI sequences. Five different MRI sequences were used to train a 3D DenseNet-121 model. At test time, it classified the series type of an input 3D volume.}
\label{fig_money}
\end{figure}

An automated approach to classify the type of MRI sequence would be beneficial and reduce the oversight needed by radiologists while improving their efficiency. Prior work in this area has mostly focused on classifying the series type in brain MRI studies \cite{Liang2021_metadataRF,ranjbar,Mello2021_ResNet18,noguchi}. Mello et al. \cite{Mello2021_ResNet18} achieved a classification accuracy of 99.27\% with the ResNet-18 architefcture trained on 3D brain MRI volumes. MRI series along with DICOM header information was used by Liang et al. \cite{Liang2021_metadataRF} to achieve near-perfect classification metrics for brain MRI sequence identification. However, this approach was still reliant on DICOM header data for proper categorization, and this may not always be readily available. There is no previous literature for the classification of MRI sequences acquired at the level of the body. We address this gap in our pilot work, and propose an automated framework for the classification of multi-parametric MRI sequences obtained at the level of the chest, abdomen, and pelvis. The proposed algorithm focuses on differentiating between the venous phase of the DCE imaging, T2-weighted (T2w), T2 fat-suppressed (T2FS), DWI, and ADC series. We demonstrate the efficacy of our model by showing consistent $F_{1}$ metrics across three deep learning architectures (DenseNet-121, ResNet-50, and ResNet-101) applied to MRI sequences of the body and brain. 

\section{METHODS}

\subsection{Data Curation}
The Picture Archiving and Communication System (PACS) at the NIH Clinical Center was queried for patients who underwent an MRI of the chest, abdomen, and pelvis between January 2015 and September 2019. 2,231 MRI studies were collected, and each series in a study was manually verified. Sequences that were empty, erroneous, or had repeated regions were excluded (n = 359), which yielded 1872 studies from 1399 patients (ages 4 - 85 years). As the majority of the patients were imaged using three different Siemens scanners (Aera, Verio, BioGraph mMR), only the 1,703 studies (1234 patients) from these scanners that contained all 5 sequences (venous phase of the DCE sequences, T2w, T2FS, DWI, ADC) were included in the final cohort. In these studies, there were often multiple DWI sequences (minimum 1, maximum 3) acquired with low (0 - 200 $s/mm^2$), intermediate (400 - 800 $s/mm^2$), and high (800 - 1400 $s/mm^2$) b-values. For our work, we utilized all the available DWI sequences with different b-values. Data (1234 patients, 1703 studies) were randomly split on a patient level into training (70\%, 864 patients), validation (10\%, 123 patients), and test (20\%, 247 patients) splits, respectively. In this manner, all the studies associated with the same patient were assigned to the same data split. 

Each MRI series was resampled to $1.5 \times 1.5 \times 7.8$ mm, and further cropped (center-crop) or padded to a constant dimension of $256 \times 256 \times 36$ voxels. The voxel intensities of each series were normalized to the [1\%, 99\%] range \cite{Kociolek2020} to remove outliers. We also used the publicly available Brain Tumor Segmentation (BraTS) dataset \cite{Menze_BraTS} to test the efficacy of our model on a different anatomical region. In the BraTS dataset, pre-contrast T1, post-contrast T1-weighted, T2-weighted, and T2 Fluid Attenuated Inversion Recovery (FLAIR) sequences were available for each patient. These volumes were already resampled to consistent dimensions of $240 \times 240 \times 155$ voxels. The model was retrained on the BraTS dataset due to discrepancies in available sequence types between the BraTS and NIH datasets.

\subsection{Model}

In order to emulate prior work \cite{Mello2021_ResNet18}, we initially implemented the ResNet-50 and ResNet-101 neural networks \cite{He_ResNet} for the classification of 3D multi-parametric MRI sequences. Then, we compared the results of these models against the DenseNet-121 architecture \cite{Huang2017_DenseNet121}, which previously showed improved classification over the ResNet models. An overview of this pipeline is shown in Fig. \ref{fig_money}. After identifying the best performing model from these experiments, we sought to determine the generalization capacity of our approach to other anatomical regions. Thus, we trained and tested our best performing DenseNet-121 model on the Brain Tumor Segmentation (BraTS) dataset. 

\begin{figure}[htb]
    \centering
    \begin{minipage}{\linewidth}\centering
    
    \subcaptionbox{GT: vDCE , Pred: vDCE}{\includegraphics[width=.3\linewidth]{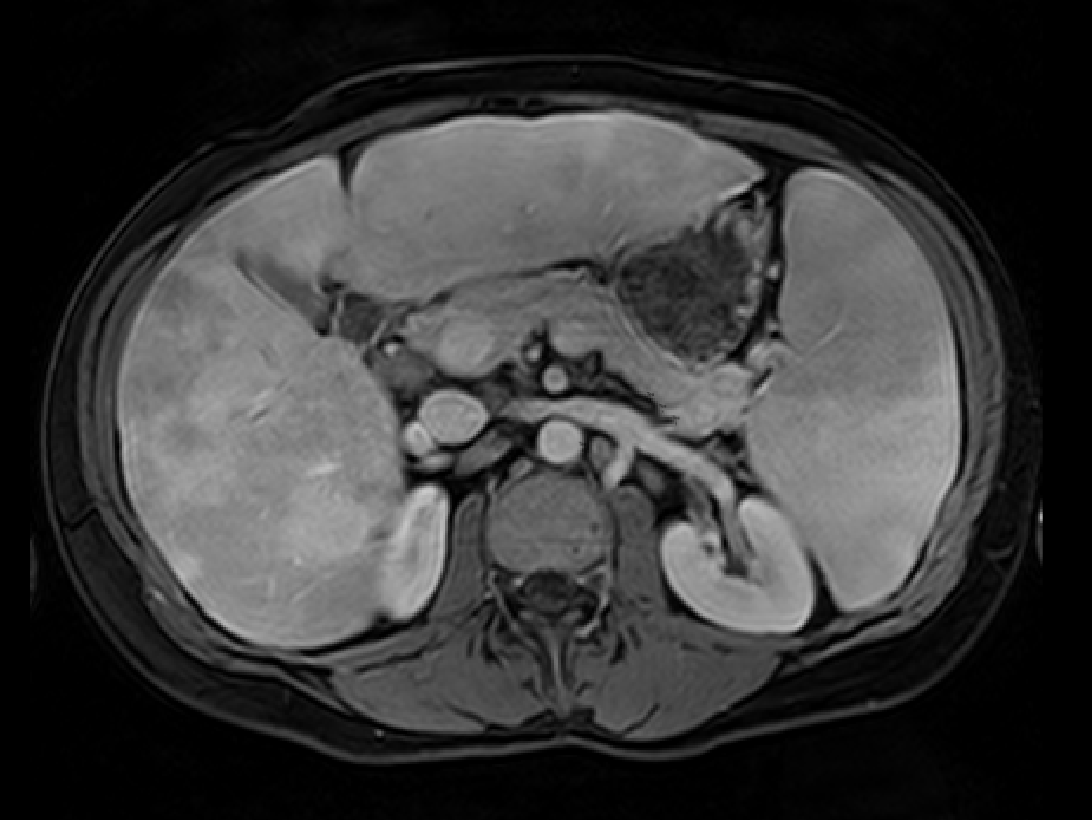}}
    \subcaptionbox{GT: T2w , Pred: T2w}{\includegraphics[width=.3\linewidth]{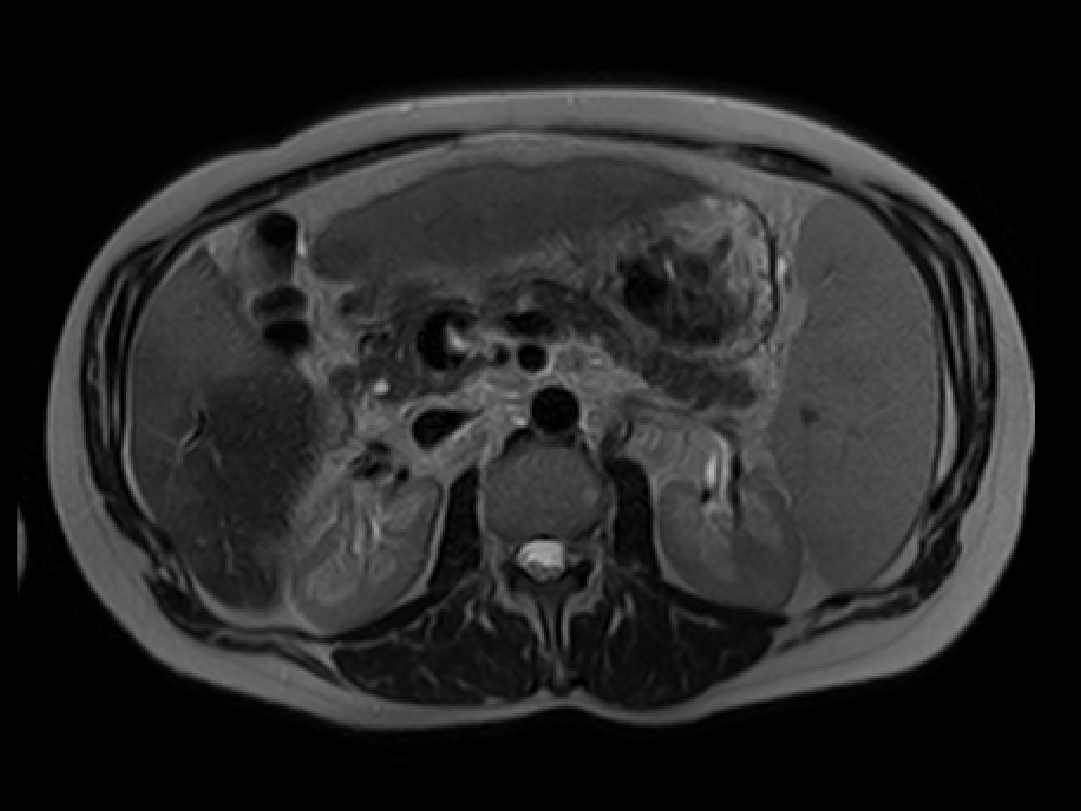}}
    \subcaptionbox{GT: ADC , Pred: ADC}{\includegraphics[width=.3\linewidth]{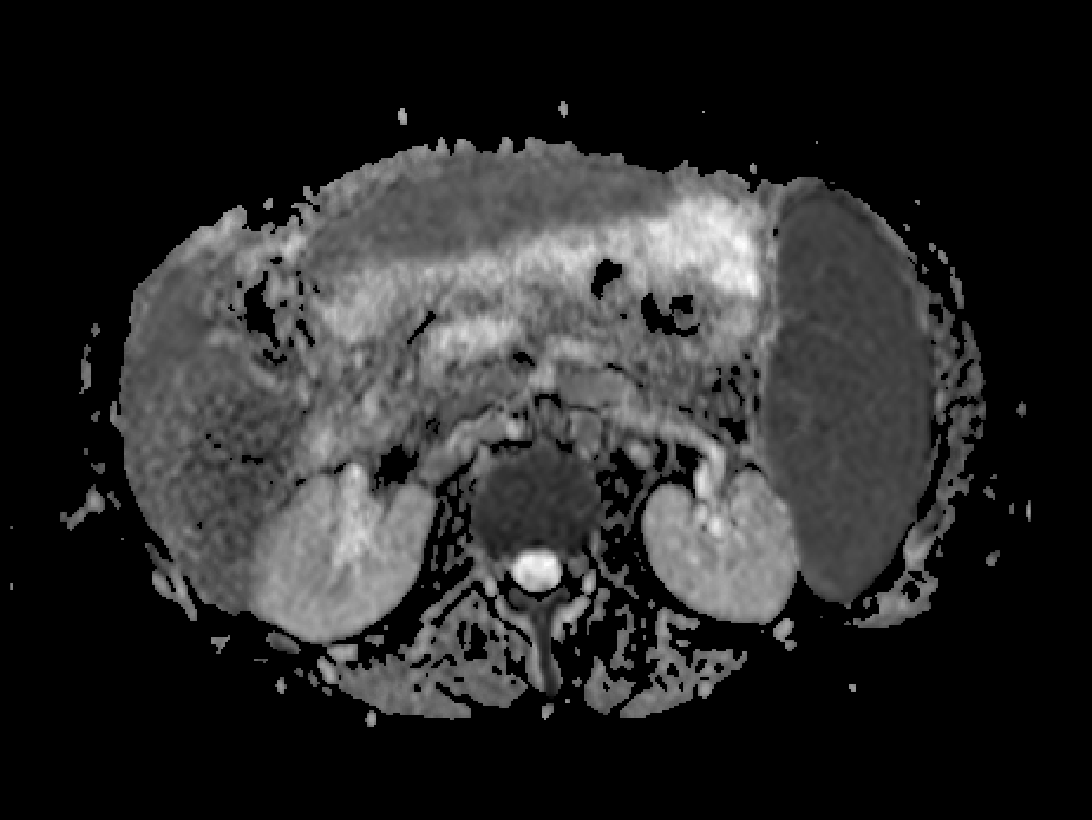}}
        
    \medskip
    
    \subcaptionbox{GT: ADC , Pred: T2w}{\includegraphics[width=.3\linewidth]{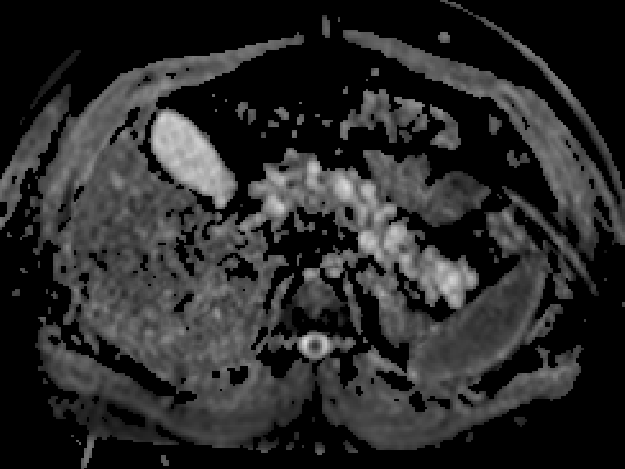}}
    \subcaptionbox{GT: T2w , Pred: DWI}{\includegraphics[width=.3\linewidth]{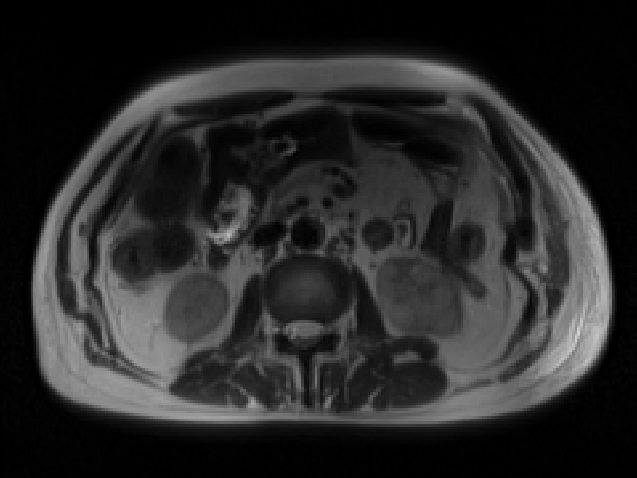}}
    \subcaptionbox{GT: DWI, Pred: T2FS}{\includegraphics[width=.3\linewidth]{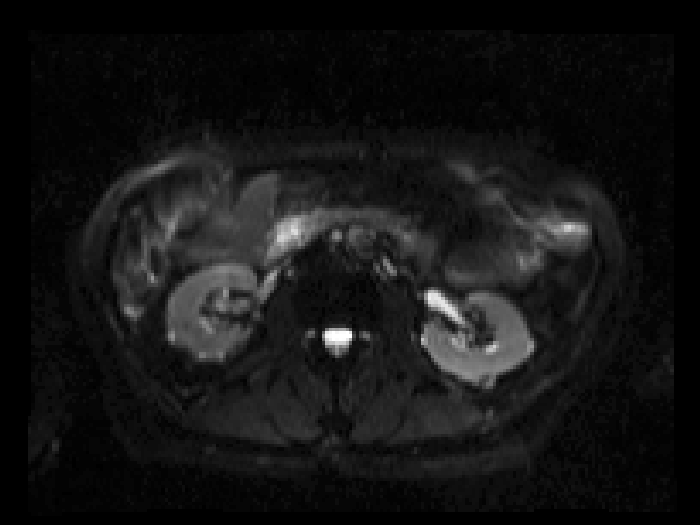}}

    \medskip

    %last two images misclassified in fold5, rest from fold1
    
    \subcaptionbox{GT: DWI, Pred: T2FS}{\includegraphics[width=.3\linewidth]{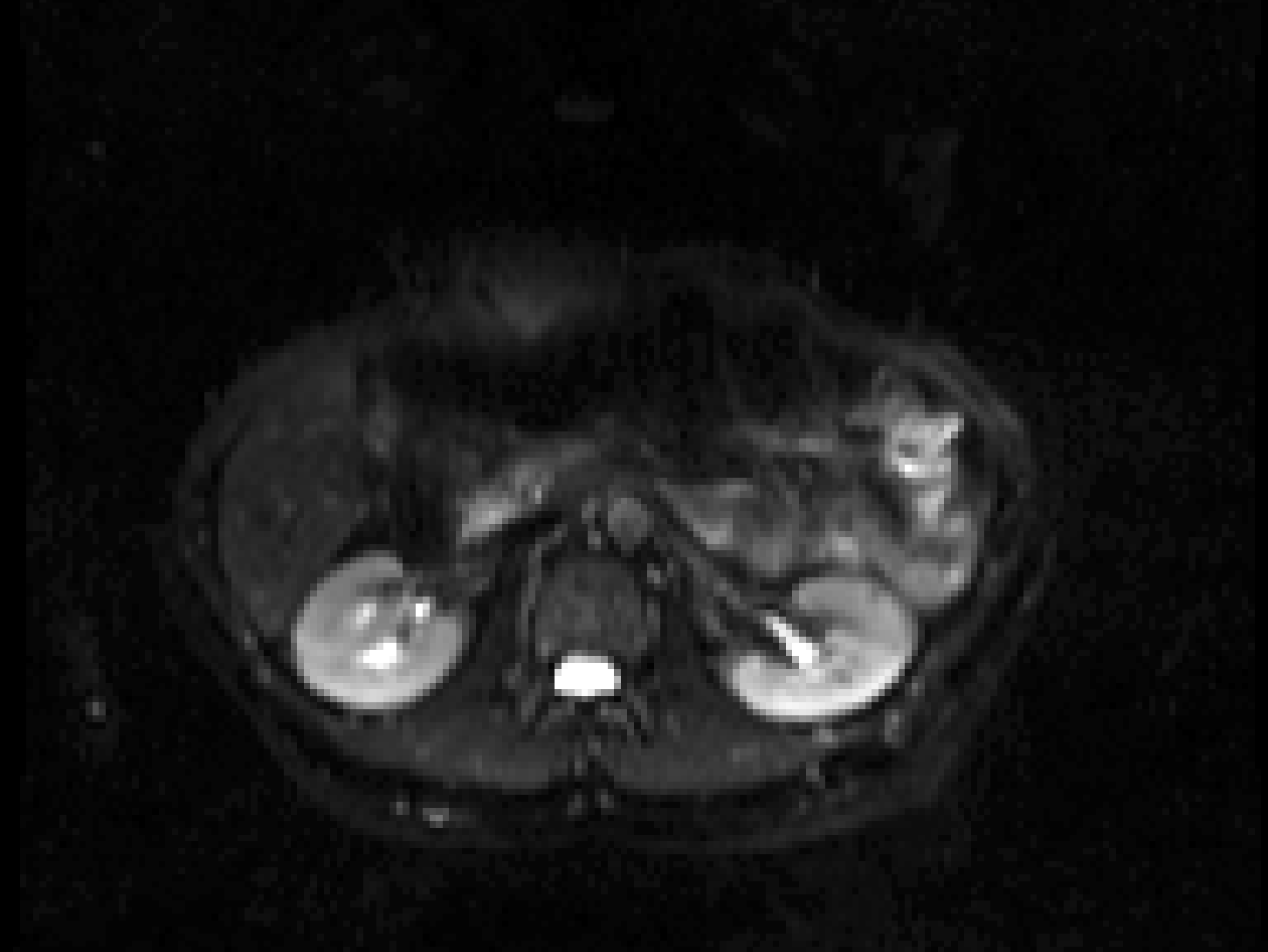}}
    \subcaptionbox{GT: ADC, Pred: T2w} {\includegraphics[width=.3\linewidth]{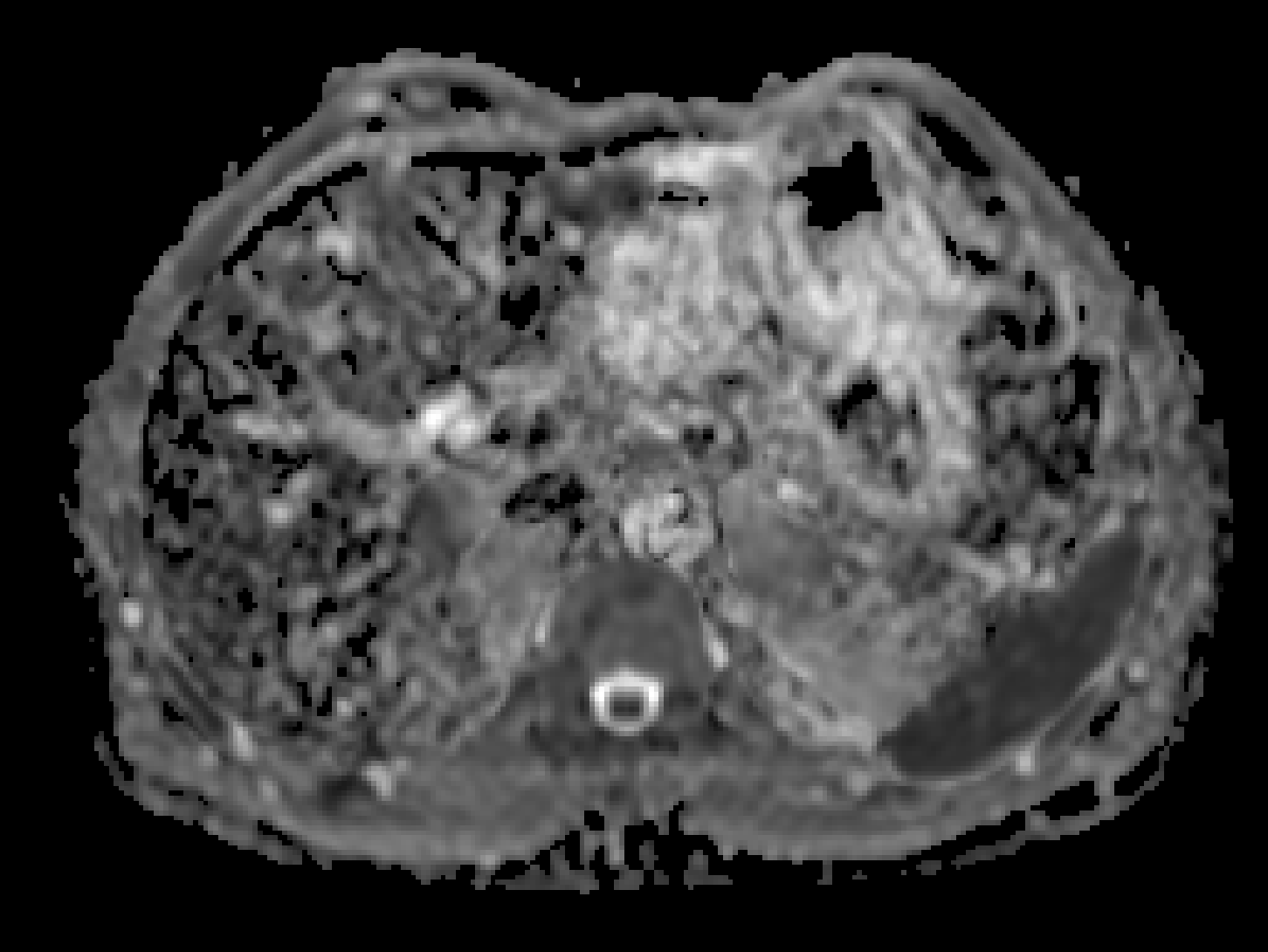}}
    \subcaptionbox{GT: T2FS, Pred: DWI}{\includegraphics[width=.3\linewidth]{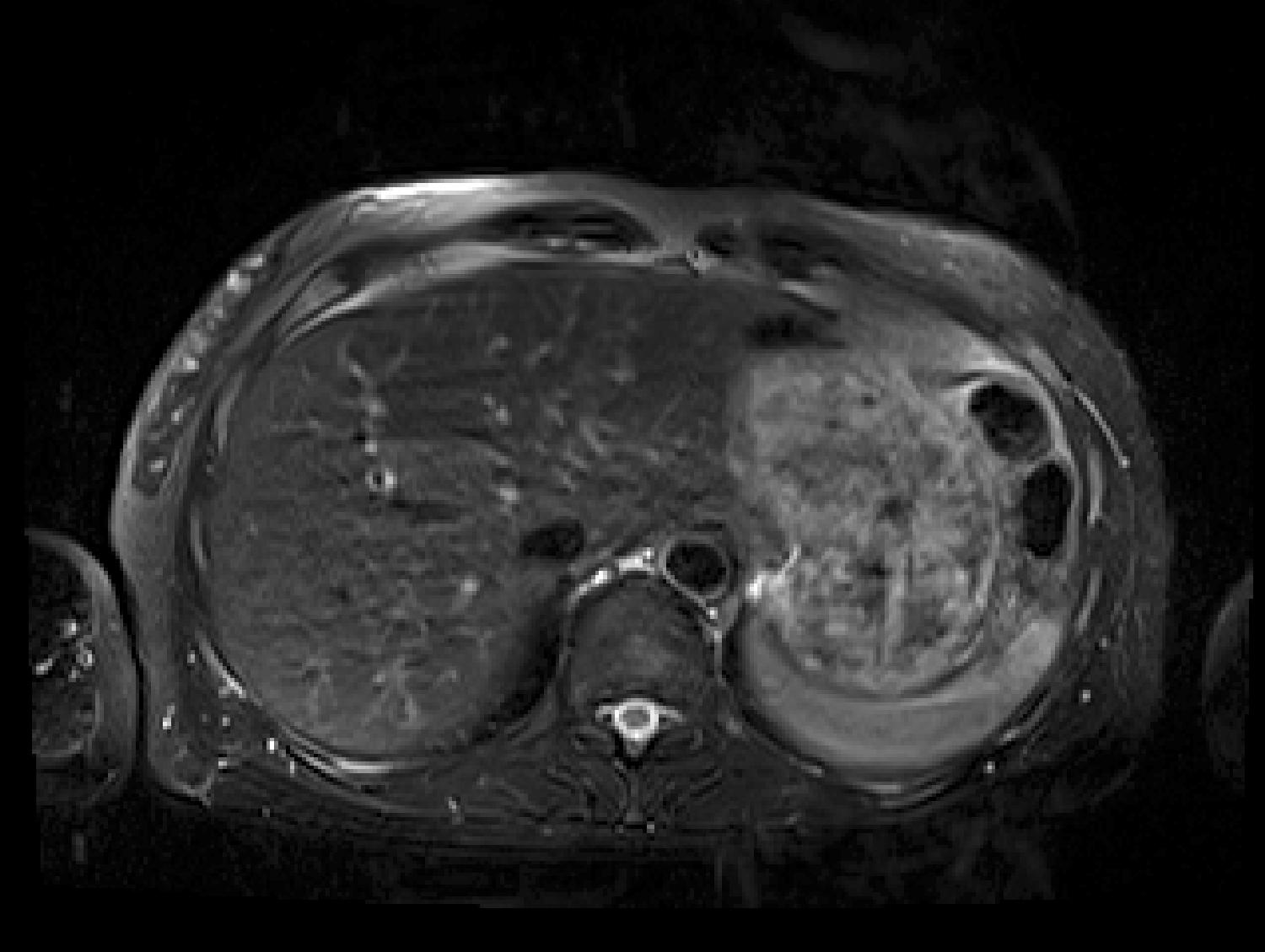}}

    \smallskip
    \caption{Qualitative results of the 3D DenseNet-121 model for correctly classified (top row) and misclassified volumes (middle and bottom rows). vDCE is an abbreviation for the venous phase of DCE imaging.}
    
    \label{fig:mri_sequence_type}
    
    \end{minipage}

\end{figure}

\section{EXPERIMENTS AND RESULTS}
\label{sec:sections}

\subsection{Implementation}

All the neural network architectures were trained with the Medical Open Network for Artificial Intelligence (MONAI) framework \cite{Cardoso_MONAI}. MONAI is an open-source machine learning framework for medical imaging built on a PyTorch base. Five-fold cross validation was then conducted with the separately held-out test split to minimize the likelihood of overfitting. Each model was trained with the cross entropy loss using a batch size of 2 and a learning rate of .0001 for a total of 25 epochs. These parameters were chosen after a grid search across the space of possible values. For each fold, the model weights associated with the epoch achieving the highest validation accuracy was saved. At inference time, ensemble metrics for each model were aggregated by averaging the accuracy, precision, recall, and $F_{1}$ score of each fold. All experiments were carried out on a NVIDIA DGX workstation running Ubuntu 20.04 LTS with a NVIDIA A100 GPU.

\subsection{Results}

A summary of results is shown in Table~\ref{table:DL_results}, and examples of correctly classified and misclassified series are shown in Fig.~\ref{fig:mri_sequence_type}. From Table \ref{table:DL_results}, the 3D DenseNet-121 model obtained the highest $F_{1}$ score of 99.5\%. The ResNet-50 and ResNet-101 models fell narrowly short by each achieving an $F_{1}$ score of 99.26\%. Our algorithm was determined to be generalizable to the brain by yielding clinically acceptable accuracy and $F_{1}$ scores of 97.83\% and 97.97\% respectively when trained and tested on the BraTS dataset. We did not perform model hyperparameter tuning while training with the BraTS dataset, and this explains the lower accuracy metrics. However, our model yielded similar results to that of Liang et al. \cite{Liang2021_metadataRF} without relying on potentially inaccurate DICOM header information. Of the 11,160 aggregate classifications across the 5-fold cross validation, 22 DWI volumes were incorrectly classified as T2FS. Conversely, only 1 T2w volume in each fold was misclassified. Similarly, volumes were misclassified as the venous phase of DCE and ADC only once for each sequence type across all folds.

%%% example of figure format is from here:
%%% https://tex.stackexchange.com/questions/311183/how-to-put-multi-subfigure-on-a-single-page-with-different-main-caption

\begin{table}[ht]
\caption{Classification results of the different networks tested on the NIH abdomen and BraTS datasets respectively. Bold font indicates best results.} 
\label{table:DL_results}
\begin{center}       
\begin{tabular}{l|c|c|c|c|c}
\hline
Architecture & Dataset & Accuracy & Precision & Recall & $F_{1}$ Score \\
\hline
DenseNet-121 & NIH Dataset & \textbf{99.50\%} & \textbf{99.51\%} & \textbf{99.50\%} & \textbf{99.50\% (99.29\%-99.71\%)} \\

ResNet-50 & NIH Dataset & 99.27\%  & 99.27\%  & 99.27\% & 99.26\% (99.01\%-99.51\%)\\

ResNet-101  & NIH Dataset & 99.27\% & 99.27\%  & 99.26\% & 99.26\% (99.01\%-99.51\%)\\

DenseNet-121  & BraTS & 97.83\% & 97.87\%  & 97.83\% & 97.83\% (97.40\%-98.26\%)\\
\hline 
\end{tabular}
\end{center}
\end{table}

\section{DISCUSSION AND CONCLUSION}

Overall, our algorithm achieved clinically acceptable classification performance for both DenseNet and ResNet architectures. Although the 5-fold ensemble metrics were marginally higher for the DenseNet implementation, the ResNet model proved to be a viable alternative that is competitive. Our proposed model minimized classification errors by relying solely on imaging data rather than potentially inaccurate DICOM header information. Despite the superior performance of our framework, several trends were observed among misclassified volumes. First, the most common error was DWI volumes incorrectly classified as T2FS. Due to the similar patterns of voxel intensity distributions across anatomical regions especially at low b-values for DWI, there may have been a perceived increase in difficulty in discerning between DWI and T2FS volumes as seen in Fig. \ref{fig:mri_sequence_type}(i). Next, T2w volumes were the least likely to be misclassified as another sequence type. Similarly, volumes had the lowest liklihood of being misclassified as the venous phase of DCE and ADC, which were each incorrectly predicted only once across all folds.

A limitation of our work is the ability of the framework to classify only five sequence types. This excludes a variety of MRI series, such as the other DCE phases (arterial, delayed) and T1 pre-contrast series. Additionally, the 3D DenseNet-121 model was solely trained on data acquired by Siemens scanners. Therefore, the generalization of the model to data obtained by scanners from other manufacturers is undetermined. Given the breadth of manufacturers in the market, this presents a barrier to implementation at other research and clinical institutions. One potential remedy to this pitfall is to add data from these domains and retrain the method to account for these differing distributions.

\section{ACKNOWLEDGEMENTS}      
 This work was supported by the Intramural Research Programs of the NIH Clinical Center. The work utilized the computational resources of the NIH HPC Biowulf cluster.

% References
\clearpage
\bibliography{REFERENCES} % bibliography data in report.bib
\bibliographystyle{spiebib} % makes bibtex use spiebib.bst

\end{document}